\documentclass[12pt]{article}

\input{epsf.sty}

\def\fo{\hbox{{1}\kern-.25em\hbox{l}}}

\def\slashchar#1{\setbox0=\hbox{$#1$}           
   \dimen0=\wd0                                 
   \setbox1=\hbox{/} \dimen1=\wd1               
   \ifdim\dimen0>\dimen1                        
      \rlap{\hbox to \dimen0{\hfil/\hfil}}      
      #1                                        
   \else                                        
      \rlap{\hbox to \dimen1{\hfil$#1$\hfil}}   
      /                                         
   \fi}                                         %

\def\beq{\begin{equation}}
\def\eeq{\end{equation}}
\def\eq{\end{equation}}
\def\to{\rightarrow}

\def\mEt{\mbox{${\hbox{$E$\kern-0.6em\lower-.1ex\hbox{/}}}_T$}\, } 
\def\hinv{H_{\rm inv}}
\def\hsm{H_{\rm SM}}

\def\nuR{\overline \nu_{R}}

\def\bsg{\ifmmode B\to X_s\gamma\else $B\to X_s\gamma$\fi}
\def\bsll{\ifmmode B\to X_s\ell^+\ell^-\else $B\to X_s\ell^+\ell^-$\fi}
\def\bstt{\ifmmode B\to X_s\tau^+\tau^-\else $B\to X_s\tau^+\tau^-$\fi}
\def\shat{\ifmmode \hat{s}\else $\hat{s}$\fi}

\newcommand{\newc}{\newcommand}

\newc{\lcal}{\int {\cal L}dt}
 
\newc{\LSP}{{\chi^0_1}}
\newc{\stauR}{{\tilde \tau_R}}
\newc{\stau}{{\tilde \tau_1}}
\newc{\mstop}{m_{\tilde{t}}}
\newc{\mHpm}{m_{H^\pm}}
\newc{\gsim}{\lower.7ex\hbox{$\;\stackrel{\textstyle>}{\sim}\;$}}
\newc{\lsim}{\lower.7ex\hbox{$\;\stackrel{\textstyle<}{\sim}\;$}}
\newc{\ie}{{\it i.e.}}          
\newc{\etal}{{\it et al.}}
\newc{\eg}{{\it e.g.}}          
\newc{\kev}{\hbox{\rm\,keV}}            
\newc{\mev}{\hbox{\rm\,MeV}}            
\newc{\gev}{\hbox{\rm\,GeV}}            
\newc{\tev}{\hbox{\rm\,TeV}}
\newc{\xpb}{\hbox{\rm\, pb}}
\newc{\xfb}{\hbox{\rm\, fb}}

\newc{\mtop}{m_t}
\newc{\mbot}{m_b}
\newc{\mz}{m_Z}
\newc{\mw}{M_W}
\newc{\alphasmz}{\alpha_s(m_Z^2)}
\newc{\swsq}{\sin^2\theta_W}
\newc{\tw}{\tan\theta_W}
\newc{\cw}{\cos\theta_W}
\newc{\sw}{\sin\theta_W}
\newc{\BR}{\hbox{\rm BR}}
\newc{\zbb}{Z\to b\bar}
\newc{\Gb}{\Gamma (Z\to b\bar b)}
\newc{\Gh}{\Gamma (Z\to \hbox{\rm hadrons})}
\newc{\rbsm}{R_b^\hbox{\rm sm}}
\newc{\rbsusy}{R_b^\hbox{\rm susy}}
\newc{\drb}{\delta R_b}

\newc{\sgn}{\mbox{sgn}}
\newc{\tbeta}{\tan\beta}
\newc{\uL}{{\tilde u_L}}
\newc{\uR}{{\tilde u_R}}
\newc{\cL}{{\tilde c_L}}
\newc{\cR}{{\tilde c_R}}
\newc{\tL}{{\tilde t_L}}
\newc{\tR}{{\tilde t_R}}
\newc{\dL}{{\tilde d_L}}
\newc{\dR}{{\tilde d_R}}
\newc{\sL}{{\tilde s_L}}
\newc{\sR}{{\tilde s_R}}
\newc{\bL}{{\tilde b_L}}
\newc{\bR}{{\tilde b_R}}
\newc{\eL}{{\tilde e_L}}
\newc{\eR}{{\tilde e_R}}
\newc{\mhp}{m_{H^\pm}}
\newc{\mhalf}{m_{1/2}}
\newc{\emt}{{e/\mu /\tau}}

\newc{\lR}{\tilde{l}_R}
\newc{\lL}{\tilde{l}_L}
\newc{\nL}{\tilde{\nu}_L}
\newc{\na}{\chi^0_1}
\newc{\nb}{\chi^0_2}
\newc{\nc}{\chi^0_3}
\newc{\nd}{\chi^0_4}
\newc{\ca}{\chi^{\pm}_1}
\newc{\cb}{\chi^{\pm}_2}
\newc{\camp}{\chi^\mp_1}
\newc{\cbmp}{\chi^\mp_1}
\newc{\capos}{\chi^{+}_1}
\newc{\caneg}{\chi^{-}_1}
\newc{\phit}{\phi_t}
\newc{\phib}{\phi_b}
\newc{\phiew}{\phi_{ew}}
\newc{\htz}{h^0_t}
\newc{\hbz}{h^0_b}
\newc{\hewz}{h^0_{ew}}
\newc{\hsmz}{h^0_{sm}}
\newc{\huz}{h^0_u}
\newc{\hsusyz}{h^0_{susy}}

\def\EPC#1#2#3{Eur. Phys. J. C {\bf #1}, #3 (19#2)}
\def\NPB#1#2#3{Nucl. Phys. B {\bf #1}, #3 (19#2)}
\def\PLB#1#2#3{Phys. Lett. B {\bf #1}, #3 (19#2)}

\def\PRD#1#2#3{Phys. Rev. D {\bf #1}, #3 (19#2)}
\def\PRL#1#2#3{Phys. Rev. Lett. {\bf#1}, #3 (19#2)}

\def\ZPC#1#2#3{Zeit. f\"ur Physik C {\bf #1}, #3 (19#2)}

\def\beq{\begin{equation}}
\def\eeq{\end{equation}}
\def\bea{\begin{eqnarray}}
\def\eea{\end{eqnarray}}
\catcode`@=11
\long\def\@caption#1[#2]#3{\par\addcontentsline{\csname
  ext@#1\endcsname}{#1}{\protect\numberline{\csname
  the#1\endcsname}{\ignorespaces #2}}\begingroup
    \small
    \@parboxrestore
    \@makecaption{\csname fnum@#1\endcsname}{\ignorespaces #3}\par
  \endgroup}
\catcode`@=12

\def\jfig#1#2#3{
 \begin{figure}
 \centering
 \epsfysize=3.0in
 \hspace*{0in}
 \epsffile{#2}
 \caption{#3}
 \label{#1}
 \end{figure}}

\begin{document}
\begin{titlepage}

\begin{flushleft}
\end{flushleft}
\begin{flushright}
hep-ph/9903259 \\
FERMILAB-PUB-99/037-T\\
CERN-TH/99-52
\end{flushright}

\vspace{1cm}

\vspace*{17.3cm}

\begin{flushleft}
CERN-TH/99-52 \\
March 1999\\
\end{flushleft}

\vspace*{-19.0cm}



\huge
\begin{center}
{\Large\bf
Motivation and detectability of an invisibly-\\
decaying Higgs boson at the Fermilab Tevatron}
\end{center}

\large

\vspace{.15in}
\begin{center}

Stephen P.~Martin${}^{a}$ and James D.~Wells${}^{b}$

\small

\vspace{.1in}
${}^{(a)}${\it Department of Physics,
Northern Illinois University, DeKalb IL 60115 {$\rm and$}\\
}
{\it Fermi National Accelerator Laboratory,
P.O. Box 500, Batavia IL 60510 \\} 
\medskip 
${}^{(b)}${\it CERN, Theory Division,
 CH-1211 Geneva 23, Switzerland \\}

\end{center}
 
\vspace{0.15in}
 
\begin{abstract}

A Higgs boson with mass below 150 GeV has a total decay width of
less than 20 MeV into accessible Standard Model states. 
This narrow width means that the usual branching
fractions for such a light Higgs boson are highly susceptible to any new
particles to which it has
unsuppressed couplings. In particular, there are many reasonable and
interesting theoretical ideas that naturally imply an invisibly-decaying
Higgs boson.  The motivations include models with light supersymmetric
neutralinos, spontaneously broken lepton number, radiatively generated
neutrino masses, additional singlet scalar(s), or right-handed neutrinos
in the extra dimensions of TeV gravity.  We discuss these approaches to
model building and their implications for Higgs boson phenomenology in
future Tevatron runs.  We find, for example, that the Tevatron with 30
fb$^{-1}$ integrated luminosity can make a 3$\sigma$ observation in the
$l^+l^- +\mEt$ channel for a 125~GeV Higgs boson that is 
produced with the same
strength as the Standard Model Higgs boson but always decays invisibly. We
also analyze the $b\bar b +\mEt$ final state signal and conclude that it
is not as sensitive, but it may assist in excluding the possibility of an
invisibly-decaying Higgs boson or enable confirmation of an observed
signal in the dilepton channel. We argue that a comprehensive Higgs search
at the Tevatron should include the possibility that the Higgs boson
decays invisibly.

\end{abstract}

\bigskip

\end{titlepage}

\baselineskip=18pt
\setcounter{footnote}{1}
\setcounter{page}{2}
\setcounter{figure}{0}
\setcounter{table}{0}
\vfill
\eject

\section{Introduction}
\bigskip

The ultimate source of electroweak symmetry breaking (EWSB) is still
mysterious.
So far, progress on solving this mystery has been confined to ruling out 
ideas rather than confirming one.
Nevertheless, an explanation exists and it is the
primary purpose of the next generation colliders to find it.

We do know what the results of EWSB must be:  the $W$ and $Z$ bosons must
get mass, and the chiral fermions must get mass.  The simplest
explanation within the Standard Model (SM) is a 
scalar $SU(2)$ doublet which couples to the vector bosons via the
covariant derivative, and to the fermions via Yukawa couplings.
After spontaneous symmetry breaking, one physical scalar degree of freedom
remains -- the Higgs boson.
Given all the other measurements that have already been made 
(gauge couplings
and masses of the gauge bosons and fermions) 
the couplings of the Higgs boson to all SM particles are fixed,
and the collider phenomenology is completely determined as a function of
only one parameter, the Higgs boson mass.

The correct theory may be much different than our simplest notion.
Low-energy supersymmetry, for example, is a rather mild deviation from
the Standard Model EWSB idea.  Nevertheless, supersymmetry 
requires at least two Higgs doublets
that contribute to EWSB and complicate
the phenomenology by having more parameters and 
more physical states in the spectrum.  Furthermore, some theories,
including supersymmetry,
may allow  other states with substantial couplings to exist which
are light enough for a Higgs boson to decay into.  
EWSB burden sharing,
or Higgs boson 
interactions with other light states are in principle just as likely
as the SM solution to EWSB.  

In this letter we would like to add to the discussion of 
EWSB possibilities
at a high luminosity Tevatron collider, by considering a Higgs
boson which decays invisibly.  The SM Higgs boson
case has been studied in great detail recently~\cite{haber}
for $\sqrt{s}=2\tev$ with high luminosity, and the prospects
for discovering a light Higgs boson ($m_h \leq 130\gev$) 
with $30\xfb^{-1}$ are promising~\cite{haber}.  
Reaching beyond $130\gev$ will be
more of a challenge, but studies in this direction appear
tantalizing~\cite{baer,han}.  
For an invisibly-decaying Higgs boson,
no studies have been performed to our knowledge.  However, we believe
it is interesting for many reasons.  

The reason non-SM Higgs phenomena is especially relevant
for the Tevatron
is because at the Tevatron a Higgs boson
is copiously produced only if its mass is less than 150 GeV or so. 
Such a light SM Higgs boson couples only
very weakly to all on-shell decay-mode states, and has a narrow
decay width in this range \cite{spira}.  For example,
$h\to f\bar f$ decays depend on the squared coupling
$m_f^2/v^2$, where $v=175\gev$. The largest mass fermion for a light
Higgs boson to decay into is the $b$ quark with $m_b\simeq 4.5\gev$,
leading to a squared coupling of order $m_b^2/v^2\lsim 10^{-3}$.
As $m_H$ is increased above 135 GeV, the decays $H\rightarrow WW^{(*)}$
begin to become more important than $H\rightarrow b\overline b$. However,
even for $m_H = (140, 150)$ GeV, the total width of a SM Higgs boson is
only about $(8, 17)$ MeV. 
Therefore, if
the light Higgs boson interacts with any new 
particle(s) in addition
to the SM particles, the resulting impact on Higgs boson decay branching
fractions could be dramatic.  For example,
an ${\cal O}(1)$
coupling of the Higgs boson to other light particles means that the Higgs
boson will decay to these new states essentially $100\%$ of the time.
If the new states happen to not be detectable, none of the standard
analyses for Higgs boson discovery would directly
apply.\footnote{In contrast, a heavy Higgs boson which is 
near or above the $WW$
threshold has a guaranteed decay mode with electroweak-strength coupling,
and
other unsuppressed decay modes enter at the $ZZ$
and $t\bar t$ thresholds.  
Therefore any new states which the Higgs boson may be
allowed to decay into will likely not completely overwhelm the SM
decay modes, so standard analyses will still be relevant
for
discovery at post-Tevatron colliders.}
Therefore, a comprehensive assessment of
EWSB  phenomenology at the Tevatron 
must include considering the possibility of a 
Higgs boson decaying invisibly.

\section{Motivation}
\bigskip

It follows from the above discussion that any
theoretical idea which allows the light Higgs boson to interact with
light invisible particles with ${\cal O}(1)$ couplings will
result in $B(H\to {\rm invisible})\simeq 100\%$.  Many possibilities
for this exist.  In the following paragraphs we list a small subset
of interesting theoretical ideas which could lead to a SM-like Higgs
boson that decays invisibly.

\bigskip
\noindent
{\it Higgs boson decays to neutralinos}
\medskip

As a first example, the lightest supersymmetric partner
(LSP), $\chi$, in supersymmetry
may be a small mixture of higgsino and bino (superpartners of the
Higgs boson and hypercharge gauge boson), and so decays of
the lightest Higgs boson into LSPs, $h\to \chi\chi$,
may have sizeable probability.  
Or, the LSP might be very nearly degenerate with other charged states
which the Higgs boson decays into, and so 
decay products of the charged states
are too soft to detect.  
If R-parity is conserved, the $\chi$ does not decay and
escapes detection.  Therefore, the Higgs boson is invisible. 
This possibility,
however, is almost excluded for minimal supersymmetry based upon gauge
coupling unification, gaugino mass unification,  and scalar mass
universality.  In this case, the lightest neutralino is mostly a
bino, and has mass approximately half that of the lightest chargino.
The present bounds
on the chargino are above about $90\gev$~\cite{neutralinos}, 
which in turn implies that
the the mass of the neutralino is at least $45\gev$ or so in minimal
supersymmetry.  
Although it is possible to
still have $h\to \chi\chi$, the parameter space remaining 
for such decays has
decreased and may continue to decrease if there is no discovery
as the CERN LEP II $e^+e^-$ collider runs proceed.
Furthermore, in minimal
supergravity parameter space the
coupling of $\chi \chi h$ is often not significantly above that
of $\bar bbh$~\cite{kane},
and so $B(h\to \chi\chi)\gg B(h\to \bar b b)$ is not necessarily expected.

However, as we stray from the naive assumptions of gaugino mass
unification and scalar mass universality, $h\to\chi\chi$ is
not as constrained by searches of chargino pair production at LEP~II.
Then the motivation is strengthened to consider the case where
this branching ratio is high, leading to an invisibly-decaying light
Higgs boson.  In supersymmetry with minimal particle content, 
the lightest Higgs boson is not
expected to be above $125\gev$~\cite{matchev}.  We shall see
later that the invisible Higgs boson can be probed with
3$\sigma$ significance up to about
$125\gev$ at the Tevatron with $30\xfb^{-1}$.  

\bigskip
\noindent
{\it Higgs boson decays to neutrinos in extra dimensions}
\smallskip

Another interesting motivation is related to neutrino mass generation in  
theories with extra dimensions opening 
up at the TeV scale~\cite{nima1,antoniadis1,dienes1}.  
In this approach, which we will call ``TeV gravity", 
no fundamental mass scale  in 
field theory should exist above a few TeV. Therefore, 
electroweak symmetry breaking,
fermion masses, flavor dynamics, and neutrino masses all must
occur near the TeV scale.  
The standard approach to neutrino mass generation
is to introduce a right-handed neutrino, and to
apply a see-saw between a heavy Majorana mass of the
neutrino ($m_M$) and a rather light Dirac mass ($m_D$).  
The lightest
eigenvalue is then $m_\nu = m_D^2/m_M$.  Typically, models prefer 
$m_M\gsim 10^{12}\gev$ either because of naturalness, or 
some considerations
in $SO(10)$ model building, etc.  In TeV gravity such high mass scales
are not available.

It is mainly theoretical prejudice that has paradoxically
made us consider extremely high mass scales to explain such low scales.
One should not forget that there are many orders of
magnitude between the neutrino mass and the weak scale in which nature
could develop the right twist to explain itself.  If TeV gravity is the
correct approach to nature, then we must find the explanation and 
identify the phenomenology that can help us discern it.  How this
is related to the invisible Higgs boson will become apparent shortly.

If the right-handed neutrino is restricted to the SM 3-brane along with
the other SM particles, neutrino masses would then need to be generated
by dynamics near or below
the TeV scale.  There are viable alternatives for this, which may
even lead to Higgs boson invisible decays~\cite{pilaftsis}.
However, one is enticed to postulate that the right-handed
neutrino is free to propagate also in the extra dimensions where gravity
propagates~\cite{nima2,dienes}.  
This is natural since $\nuR$ can be interpreted as
a singlet that has no quantum numbers to restrict it to the SM brane.
In this scenario, the $\nuR$ not only has its zero mode but 
Kaluza-Klein (KK) modes $\nuR^{(i)}$ 
separated in mass by $1/R$ 
where $R$ is the linear dimension of the compact $\delta$
dimensions, determined from
\beq
M_{\rm Pl}^2 = R^\delta M_D^{2+\delta}.
\eeq
Here
$\delta$ is the number of extra dimensions, $M_{\rm Pl}$ is the
familiar Planck mass of the effective four-dimensional theory, and
$M_D\sim 1$ TeV is
the fundamental $D=4+\delta$ dimensional gravity scale.
The absence of experimental deviations from Newtonian gravity at distances
greater than a millimeter implies that $R \lsim 10^{13}$ GeV$^{-1}$.
For $\delta = 1$, this implies $M_D \gsim 10^9$ GeV, but for $\delta \geq
2$
this does not impose any constraint stronger than $M_D \gsim 1$ TeV.

Suppose that Dirac neutrino masses arise from Yukawa couplings 
$y_\nu H\nuR \nu_L$, 
so that $m_\nu = y_\nu v$ where $v = 175$ GeV is the Higgs
vacuum expectation value.
Although the decay of $H$ to any given final state $\nu_L\nuR^{(i)}$
is proportional to
$y^2_\nu$ and extremely
small, the multiplicity of gauge-singlet right-handed neutrino KK states
below $m_H$ can be very large. It is proportional to the volume 
$R^\delta$ of the
$\delta$-dimensional space, with a momentum-space factor of order
$m_H^\delta$: 
\beq
\sum_i \rightarrow  (m_H R)^\delta\, .
\eeq 
The total partial width of $H$ into KK excitations involving neutrinos
is then of order:
\beq
\sum_i \Gamma(H \rightarrow \nu_L\nuR^{(i)}) \sim
{m_H \over 16 \pi} y_\nu^2 (m_H R)^\delta .
\eeq
Therefore the ratio of $\sum_i B(H\to \nu_L \nuR^{(i)})$
to $B(H\to b\bar b)$ can be estimated to be roughly
\beq
x \equiv \frac{\sum_i B(H\to \nu_L \nuR^{(i)})}{B(H\to \bar bb)}\simeq
{m^2_{\nu} \over 3 m_b^2}
\left ({m_H \over M_D} \right )^\delta 
\left ({M_{\rm Pl} \over M_D} \right )^2
\eeq
Now for $\delta =1$, the aforementioned constraint $M_D \gsim 10^9$ GeV
tells us that $x$ is negligibly small. 
For $\delta \geq 2$, there is no corresponding relevant constraint 
on $M_D$ and one can
estimate
\beq
x \simeq 10^{11 - \delta}
\left ({m_\nu \over 1\,{\rm eV} }\right )^2
\left ({m_H \over 100\,{\rm GeV} } \right )^\delta
\left ({1\, {\rm TeV} \over M_D} \right )^{2+\delta} \> .
\eeq
The case $\delta=2$ may run
into difficulties with nucleosynthesis, 
but for $\delta = 3$, the decays into invisible states can 
dominate~\cite{nima2}.
For example, with $m_H\gsim 100\gev$ one
can have $x \gsim 100$ even for 
$m^2_\nu = 10^{-6}$ eV$^2$ and $M_D = 1$ TeV, or for 
$m^2_\nu = 10^{-1}$ eV$^2$ and $M_D = 10$ TeV. Larger values of
$\delta$ can also give dominant invisible decays, although
the estimate is increasingly sensitive to $M_D$.
In any case, there is a strong possibility that the Higgs to 
KK neutrinos partial
width may greatly exceed
the partial widths into SM states. Note also that
there are no additional Higgs bosons necessary in this framework, allowing
$\sigma(HZ)$ to occur at the same rate as $\sigma(H_{\rm SM}Z)$ in
the SM.

\bigskip
\noindent
{\it Higgs boson decays to Majorons}
\medskip

Another approach is to assume that 
the traditional see-saw mechanism applies with a Majorana mass scale not
much larger than 1 TeV. 
In
this case, $m_M$ cannot
be much bigger than about $1\tev$.  If $m_D\sim 1\, {\rm MeV}$, and
$m_M\sim 1\tev$, then the neutrino mass is naturally 
$m_\nu \sim 1\, {\rm eV}$.
We leave it to model builders to decide why the Dirac mass of neutrinos
may be near or below about $1\mev$.  However, 
we remark that this is approximately
the electron mass and so there is precedence in nature for a Dirac
mass of a SM field to be near $1\mev$.  That is, 
no extraordinary mass scales are required in the see-saw
numerology of neutrino masses. 

The question then centers on the origin of the Majorana mass.  For us,
the important consideration is whether the Majorana mass results
from a spontaneously broken global symmetry.  If $\eta$ is a singlet scalar
field charged under a global lepton number, and if $\eta$ couples to
the neutrinos via the operator (in 2-component Weyl fermion
notation) $\lambda \eta \nuR\nuR$, then
a vacuum expectation value of $\eta$ will spontaneously break the global
lepton number and generate a Majorana mass equal to 
$\lambda\langle \eta\rangle$.  We can then identify $J={\rm Im}\, \eta$
as the Nambu-Goldstone boson of the symmetry 
breaking~\cite{chikashige,shrock}. It is easy to
write down a potential between the SM Higgs doublet $\phi$ and 
the singlet scalar $\eta$, and
to construct the interactions among mass 
eigenstates~\cite{joshipura1,joshipura2}.  The two CP-even mass eigenstates
are
\bea
H & = & \cos\theta {\rm Re}\, \phi^0 -\sin\theta {\rm Re}\, \eta \\
S & = & \sin\theta {\rm Re}\, \phi^0 +\cos\theta {\rm Re}\, \eta .
\eea
The partial widths of $H\to JJ$ and $H\to b\bar b$ can be calculated
in an arbitrary potential $V=V(\phi^\dagger \phi, \eta^\dagger\eta)$
consistent with gauge invariance and global lepton number invariance.
The ratio of these partial widths~\cite{joshipura1} (branching fractions) 
can then be expressed as
\beq
\label{br}
x \equiv \frac{B(H\to JJ)}{B(H\to \bar bb)}\simeq
 \frac{\tan^2\theta}{12}\left( \frac{m_H}{m_b}\right)^2 
 \frac{\langle \phi\rangle^2}{\langle \eta\rangle^2}.
\eeq

There are several consequences to notice from Eq.~(\ref{br}).  First, if 
$\langle \eta \rangle \gg \langle \phi\rangle$, or equivalently,
if $m_M\gg m_Z$ the Higgs boson decays into $JJ$ would not happen very often.
In the usual discussion of the Majoron model approach to neutrino masses
the prospect of $m_M\sim m_Z$ is just one possibility over a very wide
range of choices for $m_M$. However, in TeV gravity, for example, it is
required that
$m_M$ cannot be higher than the weak scale, leading to a potentially
large branching fraction of $H\to JJ$.  The second point to
notice in Eq.~(\ref{br}) is implicit.  If $\tan\theta \gg 1$ then
$B(H\to JJ)\to 100\%$.  However, in this case $\sigma (HZ)$ is
proportional to $\cos^2\theta
\to 0$, because the $HZZ$ coupling scales with $\cos\theta$.
In reality the invisible Higgs rate  in this model
for $m_Z \lsim m_H \lsim 150\gev$ 
is
\beq
{\sigma(ZH\to Z+JJ)
\over
\sigma(ZH_{\rm SM})}
= \xi(x,\cos\theta)\frac{x}{1+x}\cos^2\theta \, ,
\eeq
where $\xi\simeq 1$ represents a small correction
from $H\to WW^*,\tau\tau$ decays for 
$m_H\lsim 130\gev$, and a more sizeable $\xi\lsim 1$ correction
for larger Higgs boson mass values.

Therefore, it is impossible in this approach to 
have $\sigma(Z+JJ)>\sigma(ZH_{\rm SM})$.  Nevertheless, it is quite
possible and natural for $\sigma(Z+H\to Z+JJ)$ to be the dominant
production and decay mode of $H$, and to have a production cross-section
close to the value of a SM Higgs boson of the same mass.

\bigskip
\noindent
{\it Standard Model with an extra singlet} 
\medskip

There are many variations on the above themes which will have impact on
the production rate of the relevant Higgs boson and its decays into
invisible particles.  Rather than trying to parameterize 
all the possibilities
with complicated formulae, we instead choose to study an equally motivated
but simpler model
such that one can scale the results to any other particular idea.  In this
model there exists one gauge-singlet scalar boson and one doublet Higgs
boson whose vacuum expectation value constitutes all of 
EWSB symmetry breaking,
and which therefore couples to the $W$ and $Z$ bosons with the same
strength as the SM Higgs boson.  This minimal extension, as we will
see, has
a strong impact on the invisible width of the Higgs 
boson~\cite{datta,binoth}.

When one adds a SM singlet to the
spectrum, the full lagrangian becomes
\beq
\label{S lagrangian}
{\cal L}=
{\cal L}_{\rm SM}-m^2_S |S|^2 - \lambda' |S|^2 |H|^2-\lambda'' |S|^4,
\eeq
where $H$ is the SM doublet Higgs boson and $S=S^0+iA^0_S$ 
is the complex singlet Higgs boson. In 
writing 
Eq.~(\ref{S lagrangian}), we have
assumed 
only that $S$ is charged under a $U(1)_S$ global symmetry, and that
the Lagrangian respects this symmetry.  Without this symmetry one
could write down more terms, such as $(S^{*2}+S^2)|H|^2$, but these do not
qualitatively change the discussion below. Now if $\langle S \rangle
\not= 0$, the model is the same as the Majoron model
discussed earlier, with $U(1)_S$ playing the role of lepton number.

If $\langle S\rangle = 0$ there is no mixing between the $S$ and the
$H$, and if $m_S < m_{H}/2$ then $H\to S^0S^0,A_S^0 A_S^0$ 
are allowed to proceed
with coupling $\mu_S = 2\lambda' \langle H\rangle$.  If $\mu_S\gg m_b$ 
these decays will be near 100\% for a light Higgs boson mass
below about $150\gev$.  Since the $S^0$ does not
mix with the $H$ there will be no suppression of $ZH$ production.
Also, since $S$ has no couplings to SM gauge bosons or fermions, it
will be stable and non-interacting (invisible) in the detectors.
For the remainder of this paper we assume this model where
$\sigma(ZH)$ is unsuppressed compared to 
the SM and $H\to {\rm invisible}$ with
100\% branching fraction.  One can then scale the results to other, more
complicated models which may have suppressions in the total production
cross-section or in the invisible decay width.  One should keep in mind
that the optimal experimental analysis will be to
combine search results over all channels, including invisible
decay products,
$b\bar b$, $\tau^+\tau^-$, etc  to search for evidence in the data
of a scalar Higgs boson that may decay to several final states with
similar probabilities. 

\section{Detecting an invisibly-decaying Higgs boson with leptons}
\bigskip

The process we have found most significant in the search for an
invisibly-decaying Higgs boson in $p\overline p$ collisions at the
Tevatron
with $\sqrt{s} = 2$ TeV is 
\beq
p\bar p \to Z^*\to (Z\to l^+l^-)(\hinv\to {\rm invisible}).
\eeq
The signal is therefore two oppositely-charged same-flavor
leptons with
invariant
mass $m_Z$, accompanied by missing transverse energy from the
invisibly-decaying Higgs boson.  By $l^+l^-$ we mean $e^+e^-$ or
$\mu^+\mu^-$ and not $\tau^+\tau^-$.  The $\tau^+\tau^-$ final
states may be used to gain in significance slightly, but the
uncertainties in $\tau$ identification and invariant mass resolution
leads us to ignore this final state in the present analysis.
Again, we are assuming a theory which
is identical to the SM except that a light
singlet scalar model exists that the Higgs boson can decay into.
As discussed in the previous section, this model then 
implies that $\sigma (\hinv Z)=\sigma (H_{\rm SM}Z)$ and
$B(\hinv \to {\rm invisible})\simeq 100\%$.

The most important background is $ZZ$ production where one $Z$ boson
decays leptonically and the other $Z$ boson decays into neutrinos.
Since $ZZ$ is produced by $t$-channel processes, it is expected
that the $E_T$ 
distribution of
the $Z$ bosons will be softer (lower
energy) than the $E_T$ distribution of the $Z$ boson accompanying
$Z\hinv$, $s$-channel production.  An equivalent statement at
leading order (also NLO with a jet veto) is that the missing 
transverse energy in
the $ZZ$ background will typically be smaller than the missing
energy distribution in $Z\hinv$ events for Higgs bosons with mass near
$m_Z$. 

The next most significant background is from $W^+W^-$ production
with each $W$ decaying leptonically. (We have included here contributions
from $W \rightarrow \tau \nu$ followed by a leptonic $\tau$ decay.)
This background has a considerably softer transverse energy
distribution. As we will see in the plots and discussion below, the fact
that both of the leading backgrounds have softer transverse energy
profiles than the signal
allows the possibility to gain significance by choosing a high
cut on $\mEt$. Finite
detector resolution and smearing effects may
also favor choosing a 
higher $\mEt$ cut.
However, if the lower bound on $\mEt$ is chosen to be
too high, then one will simply run out of signal. Therefore some
intermediate choice of cut for $\mEt$ is required.

Other important backgrounds to consider arise from $WZ$,
$Wj$, and $Z^{(*)}\to \tau^+\tau^-\to l^+l^-+\mEt$. The 
$Z^{(*)}\to \tau^+\tau^-$
background is made completely
negligible by requiring that $m_{l^+l^-}\simeq m_Z$,
$\mEt>50\gev$, and $\cos(\phi_{l^+l^-})> -0.9$.  The angle
$\phi_{l^+l^-}$ is the angle between the two leptons in
the transverse plane.  The $WZ$ background requires that
a lepton from $W\to l\nu$ is not detected.  This has a rather
low probability, and our analysis requires that
the pseudo-rapidity of the missed lepton be above $|\eta|>2$.
The $Wj$ background can mimic the signal final state
if the jet registers in the detector as a lepton of
the right flavor and charge to partner with the lepton from $W\to l\nu$.
We liberally put this fake rate of $j\to l$ at $10^{-4}$.
Other backgrounds from grossly mismeasured jet energies, 
$WZ$ production with $W\rightarrow \tau\nu$, and
$t\overline t$ production 
can be
eliminated by vetoing events with a jet with
transverse energy
greater than 10 GeV and $|\eta| < 2.5$. 

We now summarize all the kinematic cuts applied in this 
analysis:
\bea
 & p_T(l^+),p_T(l^-) > 12\gev & \\
 & |\eta(l^+)|<2,~~ |\eta(l^-)|<2 & \\
 & |m_{l^+l^-}-m_Z| < 7\gev & \\
 & \cos(\phi_{l^+l^-})> -0.9 & \\
 & \mEt > 50\gev. & 
\eea
The actual analysis of signals and backgrounds was carried out at the
parton level using the CompHEP program \cite{comphep}, except for
the $WW \rightarrow \tau\ell\nu\overline\nu \rightarrow \ell\ell + \mEt$,
which
was included using the ISAJET \cite{isajet} Monte Carlo program.
We also summarize some relevant detector parameters that we assume:
\bea
 & {\rm Probability}(j\to l) = 10^{-4} & \\
 & {\rm Lost~lepton~has~}|\eta(l)|>2 & \\
 & {\rm Dilepton~id~efficiency~in~}Z\to l^+l^- = 0.7 & \\
 & {\rm NLO~K~factor}\times {\rm jet~veto} = {\rm LO}. &
\eea
The dilepton identification rate is taken from~\cite{abe}.
The last line refers to the fact that NLO calculations of EW gauge
boson production $VV'$ and gauge boson with Higgs boson production
$VH$ has a $K$ factor of slightly less than 
$1.4$~\cite{ohnemus,mrenna} at the Tevatron.
The jet veto efficiency assuming that jets must have $p_T>10\gev$ and
$|\eta_j|<2.5$ is approximately $70\%$~\cite{abe}.
Multiplying these two numbers together gives $1.4\times 0.7 \simeq 1$,
which is what we assume for the analysis. This is equivalent to simulating
background and signal at leading order (LO). Loosening the jet veto
requirement somewhat might lead to a slightly larger significance.

\jfig{backlin}{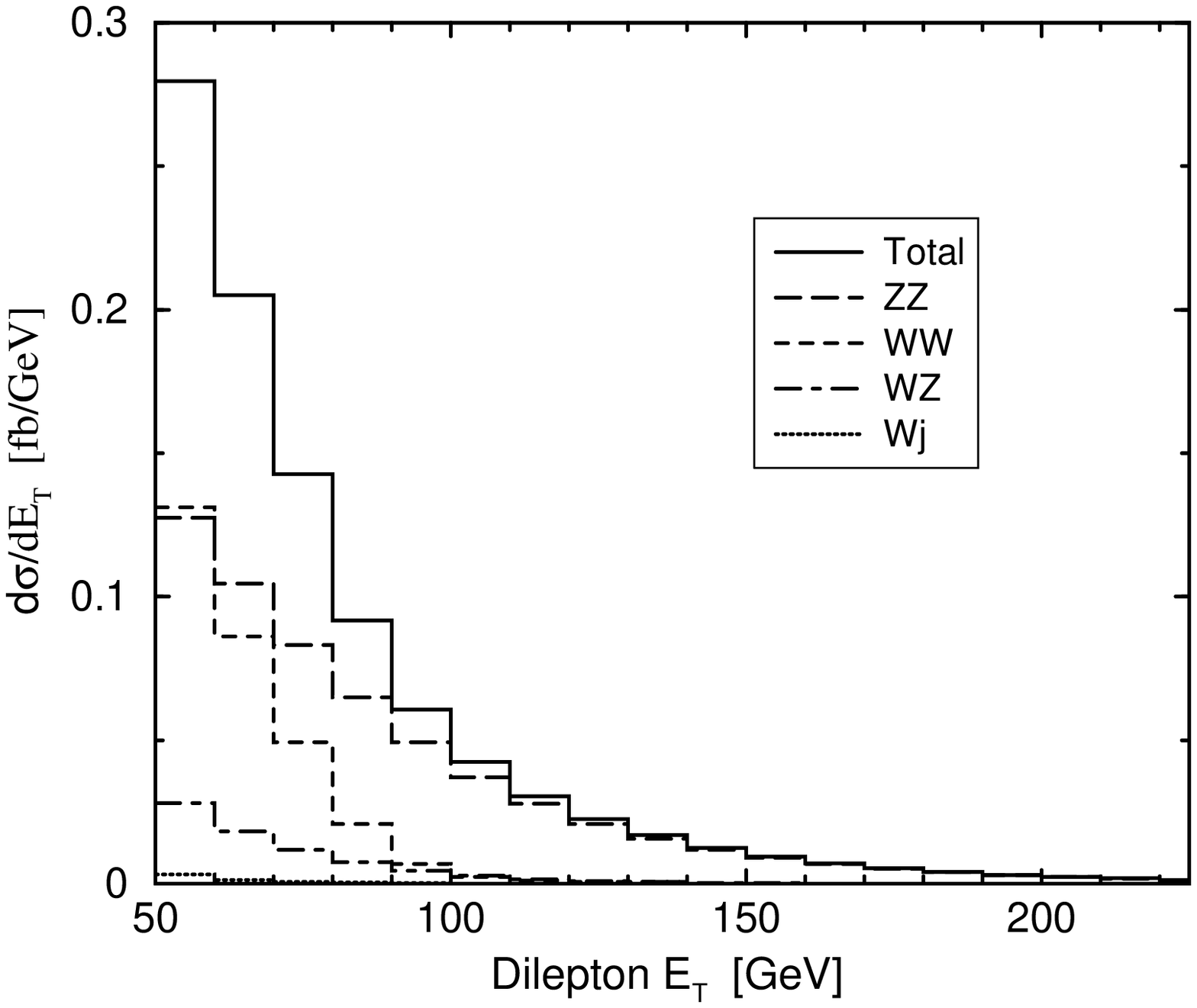}{The distribution
of the backgrounds for $l^+l^-+\mEt$ as a function of the dilepton 
$E_T$,
or equivalently, $\mEt$.  This distribution is plotted after all
cuts and efficiencies have been applied except the cut on $E_T$.}
In Figs.~\ref{backlin} and~\ref{siglin} we plot the
dilepton $E_T$ (equivalent to $\mEt$) spectrum for the background
and signal for various Higgs boson masses.
As expected, the $ZZ$ and $WW$ backgrounds are
the most significant, and the other backgrounds are down significantly
from them.  Moreover, the $WW$ background is reduced quite
significantly by choosing a higher $E_T$ cut. 
Results for the cross-sections after cuts and efficiencies are given
for the $m_H = 100$ and $130$ signals
and the total background for different choices of the $E_T$ cut, in
Table \ref{llsignal table}.
\jfig{siglin}{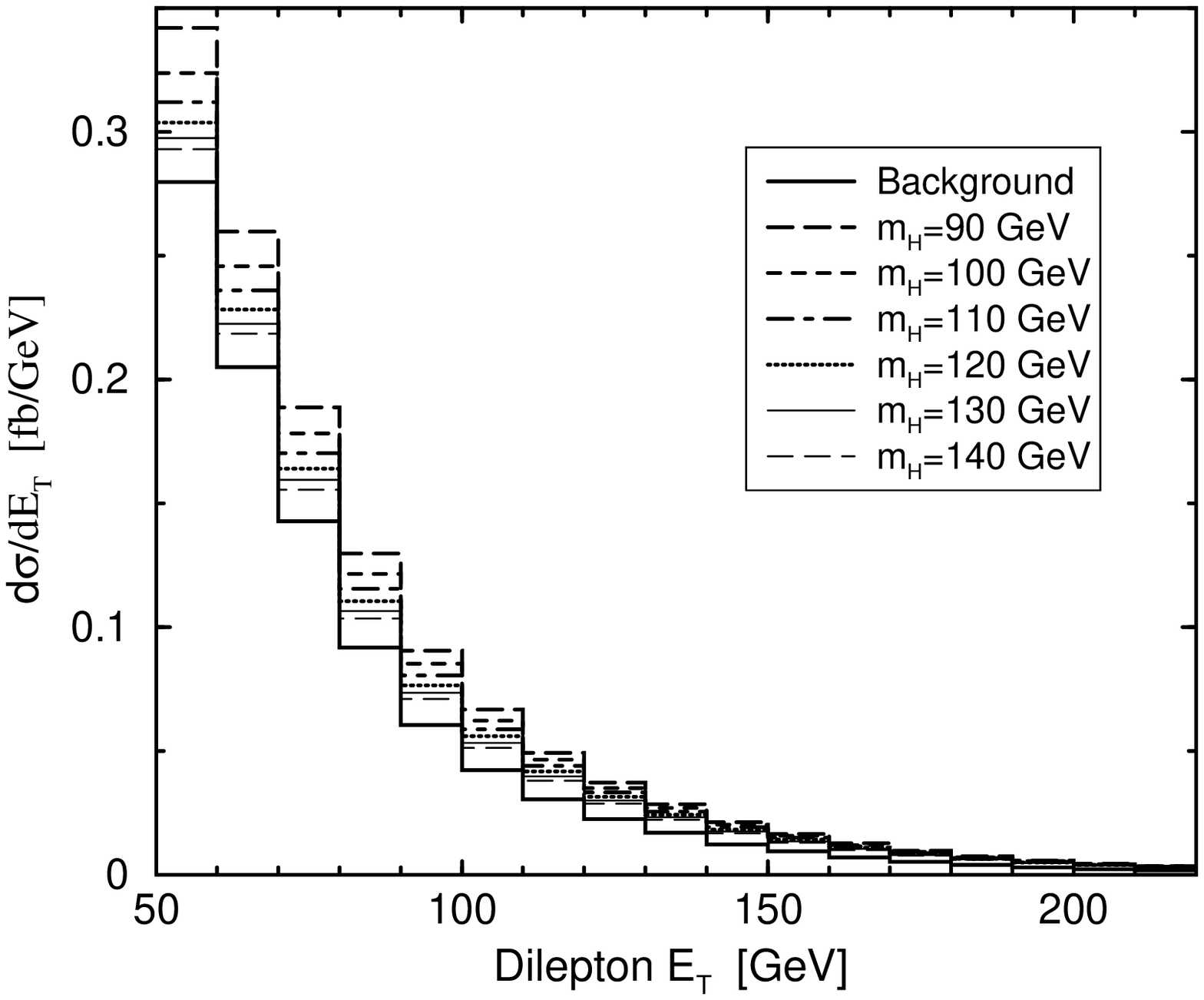}{The distribution
of the background and signal+background for $Z\hinv\to l^+l^-+\mEt$ 
as a function of the dilepton $E_T$,
or equivalently, $\mEt$.  This distribution is plotted after all
cuts and efficiencies have been applied except the cut on $E_T$.}

\begin{table}
\begin{tabular}{ccccc}
\hline\hline
$E_T$ Cut &  
$m_H = 100$ GeV signal & 
$m_H = 130$ GeV signal & 
Background \\
{}[GeV] & 
[fb] &
[fb] & 
[fb] \\
\hline
50 & 2.71 & 1.40 & 9.44 \\
60 & 2.26 & 1.22 & 6.65 \\
70 & 1.86 & 1.05 & 4.60 \\
80 & 1.50 & 0.88 & 3.17 \\
90 & 1.20 & 0.73 & 2.25 \\
100 & 0.96 & 0.60 & 1.64 \\
110 & 0.76 & 0.49 & 1.22 \\
120 & 0.60 & 0.40 & 0.91 \\
\hline\hline
\end{tabular}
\caption{Cross-sections after cuts and efficiencies for $m_H = $
100 and 130 GeV and the total background
for various choices of the $E_T$ cut. 
All cuts and efficiencies except for the $E_T$ cut have been applied.}
\label{llsignal table}
\end{table}

Using the definition
\beq
{\rm Significance} = S/\sqrt{B}
\eeq
where $S$ and $B$ are the signal and background in fb,
we plot the significance of the signal compared to background
in Fig.~\ref{significance} as a function of $E_T$.  
\jfig{significance}{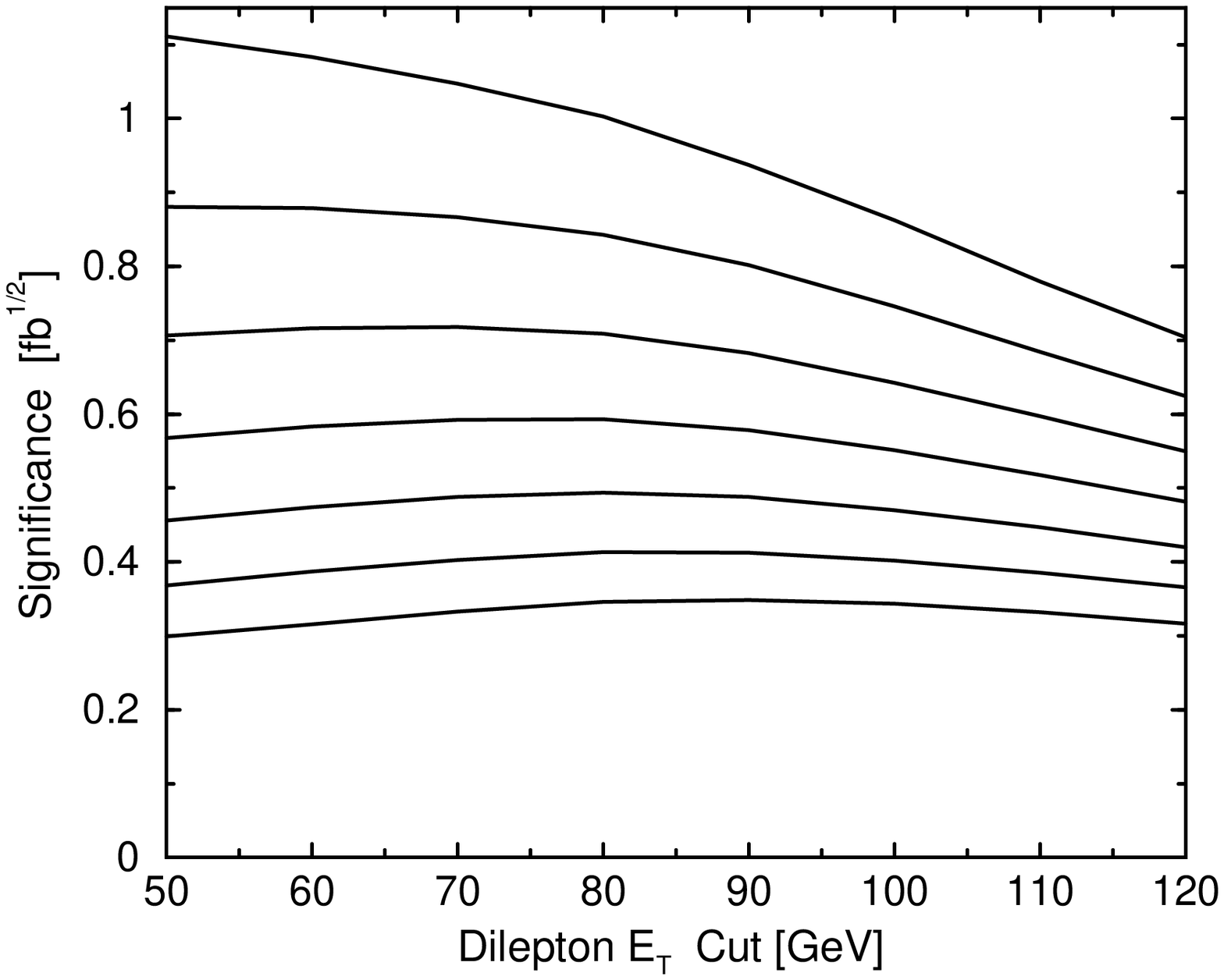}{Calculation of the significance
in fb$^{1/2}$,
defined as $S/\sqrt{B}$, as a function of the dilepton $E_T$, 
or, 
equivalently $\mEt$.  The lines from top to bottom refer to 
$m_{\hinv}=90, 100, 110, 120, 130, 140, 150$ GeV.} 
The peak of the significance curve
occurs at different $E_T$ depending on the mass of the Higgs boson.
For larger masses the significance peak is at larger $E_T$.  This is
expected
since heavier Higgs bosons will tend to carry away more missing energy
and be accompanied by more boosted $Z$
bosons, and because the $WW$
component of the background has a much softer $E_T$ distribution.  
In our analysis we choose
the $E_T$ cut for each Higgs mass in order to maximize the significance,
although the significance is a rather flat function of this cut.

We are now in position to predict how much luminosity is required
at the Tevatron to produce a $95\%$ ($1.96\sigma$)
exclusion limit, a $3\sigma$ observation,
and a $5\sigma$ discovery~\cite{conway}.
The results are shown in Fig.~\ref{luminosity} and Table~\ref{ll table}.
\jfig{luminosity}{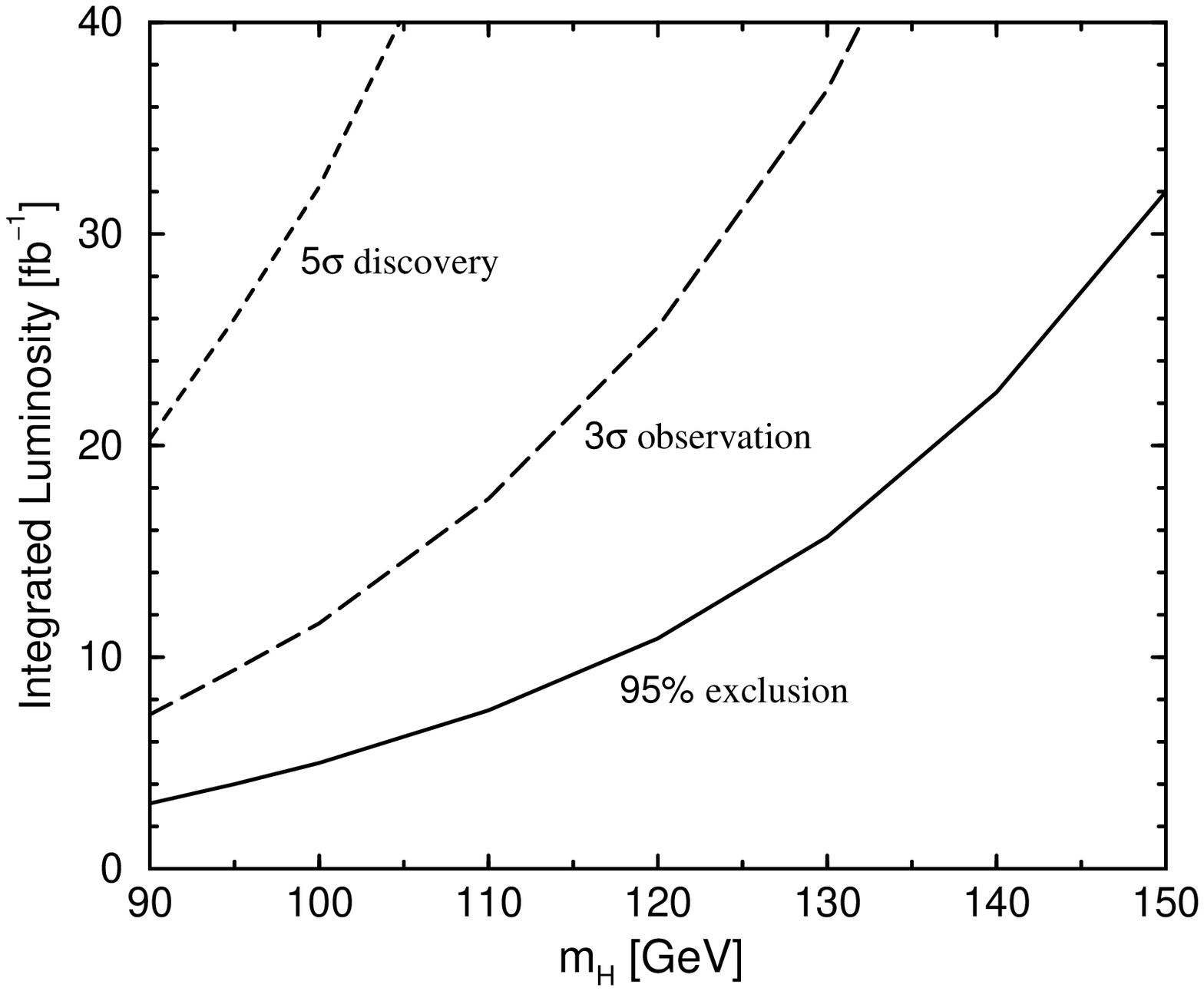}{\protect Contours of $95\%$ exclusion,
$3\sigma$ observation, and $5\sigma$ discovery in the $m_{\hinv}$ 
vs. luminosity
plane.  From this plot one can learn, for example, that with
$30\xfb^{-1}$ a $3\sigma$ observation is possible for 
$m_{\hinv}\lsim 125$ GeV.} 
\begin{table}
\begin{tabular}{ccccc}
\hline\hline
$m_{\hinv}$ [GeV]   
&  95\% Exclusion & $3\sigma$ Observation & $5\sigma$ Discovery \\
                    & Luminosity [${\rm fb}^{-1}$] &  
Luminosity [${\rm fb}^{-1}$] & Luminosity [${\rm fb}^{-1}$] \\
\hline
90  & 3.1 & 7.3 & 20 \\
100 & 5.0 & 11.6 & 32 \\
110 & 7.5 & 17.5 & 49 \\
120 & 10.9 & 26 & 71 \\
130 & 15.7 & 37 & 103 \\
140 & 23 & 53 & 146 \\
150 & 32 & 74 & 206 \\
\hline\hline
\end{tabular}
\caption{Luminosity required to make a 95\% confidence level 
exclusion, $3\sigma$
observation, and $5\sigma$ discovery of an 
invisibly decaying Higgs boson in
the $Z\hinv \to l^+l^- +\mEt$ channel.}
\label{ll table} \end{table}
If we assume that the Tevatron will accumulate a total of
$30\xfb^{-1}$ of integrated luminosity, the invisible Higgs
bosons could be excluded at the 95\% confidence level up
for $m_H$ up to nearly $150$ GeV.  
(Note, however, that the theoretical motivation for an invisibly-decaying
Higgs boson is reduced anyway as $m_H$ increases above 150 GeV and the
$H \rightarrow WW^{(*)}$ mode opens up.)
A $3\sigma$ observation is possible
for masses up to approximately $125$ GeV, and a $5\sigma$ discovery
is not possible for $m_{\hinv}>100$ GeV.  This should be compared
with LEP~II
with $\sqrt{s}=205$ GeV which should be able to discover
$\hinv$ if its mass is below $95$ GeV~\cite{campos}.
The current limit on $m_{\hinv}$ from $\sqrt{s}=184$ GeV data at LEP~II
is $80$ GeV~\cite{lep2}. 

Our results have been based only on counting events with $E_T$
larger than some cut. 
After detector responses have been more firmly established,
it may also be worth investigating whether the shape
of the $E_T$ distribution, compared to the expected background profile,
can be employed to exclude or substantiate a signal. 
In effect, the plentiful $\ell^+\ell^-+\mEt$ events with
smaller $\mEt$ (even less than 50 GeV) could be used to get a handle on
background levels which can then be tested with the higher $\mEt$ events
where the signal has its main support. This
could be done, for example, using an optimized neural net procedure. 

\section{Detecting the invisibly-decaying Higgs boson with $b$-quarks}
\medskip

Another signal that is potentially useful for discovering an
invisibly-decaying Higgs boson is\footnote{Other possible signals
involving
$p\overline p \rightarrow ZH$ followed by $Z\rightarrow jj$ without
tagged $b$-jets will suffer from large backgrounds due to 
multiple partonic contributions to $p\overline p \rightarrow jjZ
\rightarrow jj\nu\overline \nu$.}
\beq
\label{signal bb}
p\bar p \to (Z\to b\bar b)(\hinv \to {\rm invisible})=b\bar b +\mEt.
\eeq
The advantage of this signal is the increased branching
fraction of $Z\to b\bar b$ compared to $Z\to l^+l^-$. The disadvantages
are the lower efficiency for identifying $b\bar b$ final states
compared to leptonic final states, the reduced invariant
mass resolution of $Z\to b\bar b$, and more difficult background sources.  

The signal of Eq.~(\ref{signal bb}) is very similar to a $b\bar b+\mEt$
signal accessible in the SM~\cite{hedin,jesik,yao}:
\beq
\label{signal sm bb}
p\bar p\to (Z\to \nu\bar\nu)(\hsm\to b\bar b)=b\bar b+\mEt.
\eeq
Therefore, we can directly apply the background studies of
this complementary signal to the invisibly-decaying Higgs boson signal.
In
ref.~\cite{jesik}
the signal and backgrounds for $b\bar b+\mEt$ were studied using the
following
cuts and efficiency parameters:
\bea
 & p_T(b_1),p_T(b_2)>20\gev, 15\gev & \\
 & |\eta(b_{1,2})|<2 & \\
 & 
\phi(b_{1},\mEt), \phi(b_{2},\mEt) > 0.5\>{\rm radians} 
& \\
 & H_T \equiv \sum E_T(j) < 175\gev & \\
 & \mEt > 35\gev & \\
 & 70\gev < m_{bb} < 110\gev ~~({\rm loose~cut}) & \\
 & 80\gev < m_{bb} < 100\gev ~~({\rm tight~cut}) & \\
 & Z\to b\bar b~{\rm efficiency}\, = 0.49~(70\%~{\rm for~each~}b) &
\eea
The cut on $\phi(b,\mEt)$ is to ensure that the missing energy does not
originate from a grossly mismeasured $b$-jet, which may, for example, 
have neutrino(s)
carrying away much of its energy.  
We will also present results based
on the assumption of ``loose'' $m_{bb}$ invariant mass resolution,
and on the ``tight'' $m_{bb}$ invariant mass resolution as indicated
in the above.  The $b\bar b+\mEt$ total background after all cuts
are applied is $51.1\xfb$ for the ``loose'' $m_{bb}$ resolution, and
$32.3\xfb$ for the ``tight'' $m_{bb}$ resolution \cite{jesik}. These
background totals
include contributions from $ZZ$, $WZ$, $Zb\overline b$, $W
b\overline b$, single top, and $t\overline t$
production.

To apply these background studies to the present invisibly-decaying Higgs
boson situation,
we simulate the signal 
given the same kinematic cuts and efficiency
parameters.  Our simulation is at the parton level, and so we must
further take into account realistic $b$-jet energy corrections
and jet reconstruction. Also, $H_T$ is simply the sum of the two
$b$-jet
energies in our parton-level computations, but in the analysis of
ref.~\cite{hedin,jesik}
it includes a sum over other jets as well.
To take these factors into account, we can take advantage of the fact that
for $m_{H} = m_Z$, the two signals are exactly the same except for
the known effects of
branching fractions $ H,Z \rightarrow b\overline b$ and $Z\rightarrow
\ell^+\ell^-$. Therefore we normalize our total
efficiency for the
$m_{\hinv}=m_Z$ case to be equal to the efficiency found in~\cite{jesik}
for the $m_{\hsm}=m_Z$ case.  Since our signal always has
$Z\to {b\bar b}$, we can apply this overall normalization
efficiency factor for all values of $m_{\hinv}$ with little error.
A dedicated analysis of $b\bar b$ efficiencies as a function
of $\hinv$ would likely indicate a slight increase in efficiency since
the $Z$ boson $p_T$, and therefore the average $b$-jet $p_T$ values,
increases as $m_{\hinv}$ increases. 
Furthermore, the missing transverse
energy will systematically increase with $m_{\hinv}$, allowing for events
to pass the missing
energy cut
with less sensitivity to $b$-jet energy fluctuations around their intrinsic
parton values. It is quite possible that the significance can be
increased somewhat by raising the $\mEt$ cut to take advantage of this. We
therefore conclude that our approach
is justified, and perhaps yields slightly too pessimistic results.

In Table~\ref{bb table} we list the signal cross-section after cuts and
efficiencies and the
significance for the $b\bar b+\mEt$ signal.
\begin{table}
\begin{tabular}{ccccc}
\hline\hline
$m_{\hinv}$ [GeV]   &  $\sigma(b\bar b+\mEt)$ [fb]  
& $S/\sqrt{B}$ [$\sqrt{\rm fb}$] & $S/\sqrt{B}$ [$\sqrt{\rm fb}$] 
& 95\% Exclusion \\
 & & ``loose'' & ``tight'' & Luminosity [${\rm fb}^{-1}$] \\
\hline
90  & 4.10 & 0.57 & 0.72 & 7.4 \\
100 & 3.13 & 0.44 & 0.55 & 12.7 \\
110 & 2.41 & 0.34 & 0.42 & 22 \\
120 & 1.87 & 0.26 & 0.33 & 35 \\
130 & 1.46 & 0.20 & 0.26 & 57 \\ 
140 & 1.15 & 0.16 & 0.20 & 96 \\
150 & 0.91 & 0.13 & 0.16 & 150 \\
\hline\hline
\end{tabular}
\caption{Signal and significance results for the $Z\hinv \to b\bar b+\mEt$ 
process at $\sqrt{s}=2\tev$ Tevatron, estimated after cuts and
efficiencies as described in the text. The ``loose'' column refers
to assuming $70\gev <m_{bb}<110\gev$ invariant mass resolution, 
and the ``tight'' column refers to assuming $80\gev < m_{bb} < 100\gev$
invariant mass resolution.  The final column is the required luminosity
to reach a $95\%$ exclusion with the assumption of ``tight'' $bb$ invariant
mass resolution.  With $30\xfb^{-1}$, therefore, a $95\%$ exclusion could
be obtained for $m_{\hinv} \lsim 115$.}
\label{bb table}
\end{table}
The last column is the required luminosity to make a 95\% exclusion
of the invisibly-decaying Higgs boson based upon the $b\bar b+\mEt$
final state and the ``tight'' $m_{bb}$ invariant mass resolution.
With $30\xfb^{-1}$, $m_{\hinv}\lsim 115$ GeV could be excluded.
With the same luminosity,
a $3\sigma$ observation could be made for $m_{\hinv}\lsim 100$ GeV;
however, most or all of this region will likely be probed earlier by the
CERN LEP~II
$e^+e^-$ collider.
We can clearly see that at the present time
the significance of this channel in discovering the
light invisible Higgs boson is not as high as in the $l^+l^- +\mEt$
channel.  Nevertheless, $b\bar b+\mEt$ 
could be a useful channel to combine with $l^+l^- +\mEt$ to
investigate exclusion ranges, and also to obtain confirmation of
an observed signal if an excess were to develop.

\section{Conclusion}
\smallskip

In summary, there are many reasonable theoretical ideas which lead
to a light Higgs boson that most often decays invisibly.  Several of these
ideas, including Higgs decays to Majorons or right-handed neutrinos,
are made possible by mechanisms which generate neutrino masses.
Thus, our ignorance of neutrino mass generation is correlated with
our ignorance of how likely Higgs bosons will decay invisibly.
Experimentally, no theoretical prejudices should prevent the search
for this possibility.  This is especially important at the Tevatron,
since low mass Higgs bosons have very weak SM couplings, and so any
non-standard coupling of the Higgs boson to other particles
will likely garner a significant branching fraction, perhaps even
near $100\%$.

The experimental search capability of an invisible
Higgs bosons at the Tevatron requires non-SM search strategies
outlined in the previous sections.  With $30\xfb^{-1}$
one could observe (at $3\sigma$) an invisible Higgs boson with
mass up to approximately $125$ GeV in the $l^+l^- +\mEt$ channel 
and up to $100$ GeV in the $b\bar b+\mEt$ channel.  
It should be noted that the presence or absence of an excess in these
channels will require a knowledge of backgrounds which
come primarily from $ZZ$ and $WW$.
The total rates for these processes will be difficult to model with great
accuracy.
However, they can be measured directly by observation of other final
states, e.g.
$p \overline p \rightarrow ZZ
\rightarrow \ell^+\ell^- b \overline b$ 
and the rarer but clean $p \overline p \rightarrow ZZ \rightarrow
\ell^+\ell^-
\ell^{\prime +}\ell^{\prime -}$,
as well as $\ell^+\ell^-+\mEt$ events with lower $\mEt$.
The fact that these backgrounds
will need to be well-understood is a general feature 
of Higgs boson searches,
and is not strictly limited to the invisibly-decaying Higgs boson search.

The current bounds on an invisibly-decaying Higgs allow for a very
interesting window to be explored at the Tevatron. At LEP~II
with
$\sqrt{s}=205$ GeV, discovery should reach up to a mass of
at least $95$ GeV~\cite{campos}.  
At the LHC, the discovery reach may be as high
as
$150$ GeV in the gauge process 
$pp\to Z\to Z\hinv$~\cite{choudhury,frederiksen}, or
$250$ GeV in the Yukawa process $pp\to t\bar t \hinv$~\cite{gunion}.
The current published limit is $80$ GeV~\cite{lep2} from the
$\sqrt{s} = 184$ GeV run at LEP~II. 
Higgs bosons with mass much above about $150$ GeV are not likely to be
completely invisible since SM couplings to the EWSB Higgs boson exist
which are ${\cal O}(1)$ in strength, and thus lead to visible decay modes.
Therefore, an opportunity exists for a high-luminosity Tevatron
to discover or exclude the invisibly-decaying Higgs boson in the low mass
region,
which is the most likely place where an invisible Higgs boson 
would reside.

\smallskip
\noindent
{\it Acknowledgements: } We thank D.~Hedin and A.~Pilaftsis 
for helpful discussions.



\begin{thebibliography}{20}
\bibitem{pilaftsis}
For example, see
A.~Pilaftsis, \ZPC{55}{92}{275}.  Here, neutrino masses generated
radiatively below the TeV scale may lead to 
$H\to \nu_{\rm light}\bar \nu_{\rm heavy}$ invisible decays.
\end{thebibliography}
\end{document}